\documentclass[12pt,reqno,fleqn]{amsart}
\usepackage{amssymb,amsmath,amsthm,amscd,mathrsfs,amsbsy,amsfonts}
\usepackage{pstricks,pst-node}
\usepackage{graphicx}
\begin{document}

\title{Darboux transformation and solutions of the two-component Hirota-Maxwell-Bloch system*}
\author{Jieming Yang, Chuanzhong Li**,  Tiantian Li, Zhaoneng Cheng}

\dedicatory {   Department of
Mathematics,  Ningbo University, Ningbo, 315211, China }

\thanks{*Supported by the National Natural Science Foundation of China No. 11201251, the Natural Science Foundation of Zhejiang Province No. LY12A01007, the Natural Science Foundation of Ningbo City under Grant No 2013A610105, and the K. C. Wong Magna Fund in Ningbo University.\\
** Corresponding author. Email: lichuanzhong@nbu.edu.cn.}
\texttt{}

\date{}

\begin{abstract}
 In this paper, we derive n-fold Darboux transformation of the two-component Hirota and the Maxwell-Bloch(TH-MB) equations and its determinant representation. Using Darboux determinant representation, we provide soliton solutions, positon solutions of the TH-MB equations.

\end{abstract}

\maketitle

\allowdisplaybreaks
 \setcounter{section}{0}

Nonlinear science as a powerful subject  explains  all kinds of mysteries in the challenges of science and technology today.
In the nonlinear science,  solitons and positons in nonlinear dynamical systems attract a lot of
research in oceanography,  Bose-Einstein condensate,  plasma, superfluid
and especially in optics(\cite{Osborne,AB1,wuy,wuy2,wuy3}).

For a reduced dynamical equation, the erbium-doped fibre system was proven to allow soliton-type pulse propagation with pumping\cite{Nakkeeranjpa}. The Lax pair and the exact soliton
solution for Higher-order nonlinear Schr\"odinger and Maxwell-Bloch(HNLS-MB) equations with pumping was derived in \cite{Nakkeeranjpa}.
Kodama\cite{Kodama} shows that with suitable transformation and omitting the higher-order terms, higher order Nonlinear Schr\"odinger equation equation can be reduced to the Hirota equation\cite{hirotaeq}. In a similar way, after suitable choice of self steepening and self frequency effects, the                                                                                                                                                                                                                                                                                HNLS-MB equations can be reduced to a coupled system of the Hirota equation and MB equation(H-MB) which is governed by femtosecond pulse propagation through doped fibre\cite{PorsezianPRL}.

The H-MB system has been shown to be
integrable with Lax pair \cite{PorsezianPRL}. By  an efficient method which is called Darboux transformation \cite{Matveev}, the determinant representation of n-fold Darboux transformation of AKNS system was given in \cite{Hedeterminant}. In \cite{rogueHMB,HMBCPB},  Darboux transformation and the rogue wave solutions of the H-MB equations were obtained using the Darboux transformation.
Recently, n-fold Darboux transformation and positon solutions of the inhomogeneous  Hirota-Maxwell-Bloch system is revealed in \cite{chinascience}.
Soliton interactions in a generalized inhomogeneous coupled Hirota-Maxwell-Bloch system were also considered in \cite{BoTian} with symbolic computation.
The two-component  Hirota and Maxwell-Bloch equations \cite{CHMBchaos} as a coupled version of the Hirota and Maxwell-Bloch equations includes all the important effects like the SIT in a Kerr medium with magnetisation, higher order dispersion and self-steepening. These effects play a very important role in the ultra-short pulse propagation of the order of femto seconds in nonlinear optical fibres. That is why we consider the Darboux  transformation  which is very useful to derive all kinds of solutions of the TH-MB equations which should be useful in explaining phenomenons of nonlinear optics.

The two-component Hirota-Maxwell-Bloch equation is as following\cite{CHMBchaos}
\begin{eqnarray}E_{1z}& =& \frac12i\beta E_{1tt}+iA\beta E_1+\epsilon(E_{1ttt}+3AE_{1t}+3BE_1)-p_1,\\
E_{2z}& =& \frac12i\beta E_{2tt}+iA\beta E_2+\epsilon(E_{2ttt}+3AE_{2t}+3BE_2)-p_2,\\
p_{1t}& =&  -NE_1+M_{11}E_1+M_{21}E_2+\omega ip_1,\\
p_{2t}& =&  -NE_2+M_{12}E_1+M_{22}E_2+\omega ip_2,\\
M_{11t} &=& -E_1p_1^*-E_1^*p_1,\\
M_{12t} &=& -E_1^*p_2-E_2p_1^*,\\
M_{21t} &=& -E_1p_2^*-E_2^*p_1,\\
M_{22t} &=& -E_2p_2^*-E_2^*p_2,\\
N_{t} &=&E_1p_1^*+E_1^*p_1+E_2p_2^*+E_2^*p_2,\end{eqnarray}
\begin{eqnarray}A:=|E_1|^2+|E_2|^2,\ \ B:=E_{1t}E_1^*+E_{2t}E_2^*.\end{eqnarray}

The corresponding linear eigenvalue problem can take the form

\begin{eqnarray}
\Phi_t&=&U\Phi,\\
\Phi_z&=&V\Phi,
\end{eqnarray}
where
\begin{eqnarray}
U&=&\left(\begin{matrix}-i\lambda& E_1&E_2\\ -E_1^*& i\lambda&0\\ -E_2^*& 0&i\lambda
\end{matrix}\right)=-i\lambda \sigma_3+U_0,\\
V
&=&\lambda^3V_3+\lambda^2V_2+\lambda V_1+V_0+\frac{i}{\lambda+\frac{\omega}2}V_{-1},
\end{eqnarray}
\begin{eqnarray*}
V_3&:=&-8i\epsilon\left(\begin{matrix}0&0& 0\\0&1& 0\\0&0& 1
\end{matrix}\right),\ \
V_2:= \left(\begin{matrix}0&E_1&E_2\\-E_1^*&0&0\\-E_2^*&0&0\end{matrix}\right);\ \
V_1:=\left(\begin{matrix}|E_1|^2+|E_2|^2&E_{1t}&E_{2t}\\E_{1t}^*&-|E_1|^2&-E_2E_1^*\\E_{2t}^*&-E_1E_2^*&-|E_2|^2\end{matrix}\right);\\
V_0&:=& \left(\begin{matrix}E_{1t}E_1^*-E_{1t}^*E_1+E_{2t}E_2^*-E_{2t}^*E_2&E_{1tt}+2E_1(|E_1|^2+|E_2|^2)&E_{2tt}+2E_2(|E_1|^2+|E_2|^2)\\
-E_{1tt}^*-2E_1^*(|E_1|^2+|E_2|^2)&-(E_{1t}E_1^*-E_{1t}^*E_1)
&-(E_{2t}E_1^*-E_{1t}^*E_2)\\-E_{2tt}^*-2E_2^*(|E_1|^2+|E_2|^2)&-(E_{1t}E_2^*-E_{2t}^*E_1)&-(E_{2t}E_2^*-E_{2t}^*E_2)\end{matrix}\right);
\end{eqnarray*}
\begin{eqnarray}\ \ V_{-1}:=\frac12\left(\begin{matrix}N&p_1&p_2\\p_1^*&M_{11}&M_{12}\\p_2^*&M_{21}&M_{22}.
\end{matrix}\right),\ \ \sigma_3=\left(\begin{matrix}1& 0&0\\ 0&-1&0\\ 0
& 0&-1
\end{matrix}\right),\ U_0=\left(\begin{matrix}0& E_1&E_2\\ -E_1^*& 0&0\\ -E_2^*& 0&0
\end{matrix}\right).\end{eqnarray}
Here $E_1$ and $E_2$ are the slowly varying amplitudes of the signal, $p_i, i= 1,2$ and $M_{ij}, i,j= 1,2$ are the electric and magnetic polarizations respectively, $\beta$ and $\epsilon$ are constants and $N$ is the concentration of resonant atoms, $*$ means complex conjugation.

For the H-MB equation and inhomogeneous H-MB equation, the Darboux transformation is constructed in \cite{rogueHMB,chinascience}. But for the TH-MB, because of its
complexity, the Darboux transformation and reduction conditions are still not clear. Using above linear equations of TH-MB equations, one-fold Daroux transformation for TH-MB equation will be introduced in the next section.

Firstly, one consider the transformation about linear function $\Phi$ by
\begin{eqnarray}
\Phi'&=&T\Phi=(\lambda A-S)\Phi,
\end{eqnarray}
where
\begin{eqnarray}A=\left(\begin{matrix}a_{11}& a_{12}&a_{13}\\ a_{21}& a_{22}&a_{23}\\ a_{31}& a_{32}&a_{33}
\end{matrix}\right),\ \ \
S=\left(\begin{matrix}s_{11}& s_{12}&s_{13}\\ s_{21}& s_{22}&s_{23}\\ s_{31}& s_{32}&s_{33}
\end{matrix}\right).
\end{eqnarray}

New function $\Phi'$ is supposed to satisfy
\begin{eqnarray}
\Phi'_t&=&U'\Phi',\\
\Phi'_z&=&V'\Phi'.
\end{eqnarray}

Then matrix $T$ must satisfy following identities
\begin{eqnarray}\label{tequation}
T_t+TU&=&U'T,\\ \label{zequation}
T_z+TV&=&V'T.
\end{eqnarray}
One can  choose  $T=\lambda I-S$ and then the Darboux transformation on $E_1,E_2,p_1,p_2,N,M_{11},M_{12},M_{21},M_{22}$   can be got by eq. \eqref{tequation} and eq. \eqref{zequation} as following
\begin{eqnarray}U'_0=U_0+i[S,\sigma_3],\end{eqnarray}
\begin{eqnarray}\label{V'-1dressing}
V'_{-1}=(S+\frac{\omega}2) V_{-1}(S+\frac{\omega}2)^{-1},
\end{eqnarray}
and $S$ should have a condition as $s_{21}=-s^*_{12}, s_{31}=-s^*_{13}.$

We suppose
\begin{eqnarray}\label{SandH}
S=H\Lambda H^{-1}\end{eqnarray}
where
$\Lambda=\left(\begin{matrix}\lambda_1& 0&0\\ 0& \lambda_2&0\\0&0& \lambda_3
\end{matrix}\right)$, $H=\left(\begin{matrix}\Phi_1(\lambda_1,t,z)& \Phi_1(\lambda_2,t,z)& \Phi_1(\lambda_3,t,z)\\ \Phi_2(\lambda_1,t,z)& \Phi_2(\lambda_2,t,z)& \Phi_2(\lambda_3,t,z)\\ \Phi_3(\lambda_1,t,z)& \Phi_3(\lambda_2,t,z)& \Phi_3(\lambda_3,t,z)
\end{matrix}\right):=\left(\begin{matrix}\Phi_{1,1}& \Phi_{1,2}& \Phi_{1,3}\\ \Phi_{2,1}& \Phi_{2,2}& \Phi_{2,3}\\ \Phi_{3,1}& \Phi_{3,2}& \Phi_{3,3}
\end{matrix}\right).$
In order to satisfy the constraints of $S$, meanwhile make $V'_{-1}$ having similar form as $V_{-1}$, i.e. $s_{21}=-s^*_{12},s_{31}=-s^*_{13},$
following constraint will be considerd
\begin{eqnarray}\notag
\lambda_3&=&\lambda_2=\lambda_1^*,\ \
H=\left(\begin{matrix}\Phi_1(\lambda_1,t,z)& \Phi^*_2(\lambda_1,t,z)& \Phi^*_3(\lambda_1,t,z)\\ \Phi_2(\lambda_1,t,z)& -\Phi^*_1(\lambda_1,t,z)& 0\\ \Phi_3(\lambda_1,t,z)& 0& -\Phi^*_1(\lambda_1,t,z)
\end{matrix}\right).\end{eqnarray}

As the simplest Darboux transformation, the determinant representation of one-fold Darboux transformation  of the TH-MB equations will be given in the following theorem.

\emph{Theorem 1:}\label{1folddarboux}
The one-fold  Darboux transformation of the TH-MB equations is as following
\begin{eqnarray}T_1(\lambda,\lambda_1,\lambda_2,\lambda_3)=\lambda I+t_0^{[1]}=\frac{1}{\Delta_1}\left(\begin{matrix}(\mathbb{T}_1)_{11}&(\mathbb{T}_1)_{12}&(\mathbb{T}_1)_{13}\\
(\mathbb{T}_1)_{21}&(\mathbb{T}_1)_{22}&(\mathbb{T}_1)_{23}\\
(\mathbb{T}_1)_{31}&(\mathbb{T}_1)_{32}&(\mathbb{T}_1)_{33}
\end{matrix}\right),\end{eqnarray}
where for $1\leq i,j,k,m\leq 3$

\begin{eqnarray} \notag \Delta_1&=&
\det \left(W_1\right)_{3\times 3},\ (\mathbb{T}_1)_{km}=
\det \left(\begin{matrix}p_m&\lambda^{\delta_{km}}\\
W_1&q_k^T\end{matrix}\right)_{4\times 4}.\end{eqnarray}
The matrix $W_1$ and row vectors $p_m,q_k$ are defined as,
\begin{eqnarray}(W_1)_{ij}=\Phi_{j,i},\ (p_m)_j=\delta_{m,j},\ \ q_k=\left(\begin{matrix}\lambda_1\Phi_{k,1}& \lambda_2\Phi_{k,2}&\lambda_{3}\Phi_{k,3}\end{matrix}\right).\end{eqnarray}

The following identities gives Darboux transformation of the TH-MB equation,
\begin{eqnarray}\label{1darbouxgeneral}U^{[1]}=U+[\sigma_3,T_1],\ \ V^{[1]}_{-1}=T_1|_{\lambda=-\frac{\omega}2}V_{-1}T_1^{-1}|_{\lambda=-\frac{\omega}2}.\end{eqnarray}
Eqs.\eqref{1darbouxgeneral} will give one-fold Darboux transformation about $E_1,E_2$ as
\begin{eqnarray}\label{1darbouxE}E_1^{[1]}&=&E_1-2is_{12}=E_1+2i\frac{(\mathbb{T}_1)_{12}}{\Delta_1},\ \ \ E_2^{[1]}=E_2-2is_{13}=E_2+2i\frac{(\mathbb{T}_1)_{13}}{\Delta_1},\end{eqnarray}
and
one-fold Darboux transformation about $p_1,p_2,N,M_{11},M_{12},M_{21},M_{22}$ of the TH-MB equations in complicated forms.

Because the one-fold Darboux transformation about $p_1,p_2,N,M_{11},M_{12},M_{21},M_{22}$ of the TH-MB equations is in very complicated forms, therefore we have to omit it here for saving space.
This one-fold transformation will be used to generate one-soliton solution from trivial seed solution of the TH-MB equation.

In the next part, we will give determinant representation of the n-fold Darboux transformation of the TH-MB equations.
Firstly, we introduce $3n$ eigenfunctions $\Phi_i=\left(\begin{matrix}\Phi_{1,i}\\ \Phi_{2,i}\\ \Phi_{3,i}\end{matrix}\right), i=1,2,\dots,3n$
and $\Phi_i=\Phi(\lambda=\lambda_i),$
with  constraints on eigenvalues as $\lambda_{3n+1}=\lambda_{3n}=\lambda_{3n-1}^*$ and  the reduction conditions on eigenfunctions as
\begin{eqnarray}\notag
\Phi_{3n-1,2}=\Phi_{3n,3}=-\Phi_{3n-2,1}^*,\ \ \Phi_{3n-2,2}=\Phi_{3n-1,1}^*,\ \ \Phi_{3n-2,3}=\Phi_{3n,1}^*, \Phi_{3n-1,3}=\Phi_{3n,2}=0.\\
\end{eqnarray}
We can also derive the n-fold Darboux transformation in form of a huge determinant in the following theorem.

\emph{Theorem 2:}
The n-fold  Darboux transformation for solutions of the TH-MB equations can be represented as
\begin{eqnarray}\notag T_n(\lambda;\lambda_1,\lambda_2,\lambda_3,\lambda_4,\dots,\lambda_{3n})&=&\lambda^n I+t_{n-1}^{[n]}\lambda^{n-1}+\dots+t_1^{[n]}\lambda+t_0^{[n]}\\
&=&\frac{1}{\Delta_n}\left(\begin{matrix}(\mathbb{T}_n)_{11}&(\mathbb{T}_n)_{12}&(\mathbb{T}_n)_{13}\\
(\mathbb{T}_n)_{21}&(\mathbb{T}_n)_{22}&(\mathbb{T}_n)_{23}\\
(\mathbb{T}_n)_{31}&(\mathbb{T}_n)_{32}&(\mathbb{T}_n)_{33}
\end{matrix}\right),\end{eqnarray}

\begin{eqnarray}U_{0}^{[n]}=U_0+t_{n-1}^{[n]}(-i\sigma_3)-(-i\sigma_3)t_{n-1}^{[n]},\end{eqnarray}
\begin{eqnarray}\label{V-1DT}V_{-1}^{[n]}=T_n|_{\lambda=-\frac{\omega}2} V_{-1}T_n^{-1}|_{\lambda=-\frac{\omega}2},\end{eqnarray}

where for $1\leq i,j,k,m\leq 3n$

\begin{eqnarray} \notag \Delta_n&=&
\det \left(W_n\right)_{3n\times 3n},\ (\mathbb{T}_n)_{km}=
\det \left(\begin{matrix}p_m&\lambda^{n\delta_{km}}\\
W_n&q_k^T\end{matrix}\right)_{(3n+1)\times (3n+1)}.\end{eqnarray}
The matrix $W_n$ and row vectors $p_m,q_k$ are defined as,
\begin{eqnarray}(W_n)_{ij}=\Phi_{\bar j,i}\lambda_i^{[\frac j3]},\ (p_m)_j=\delta_{m,\bar j}\lambda^{\lfloor\frac j3\rfloor},\ \ q_k=\left(\begin{matrix}\lambda_1^n\Phi_{k,1}& \lambda_2^n\Phi_{k,2}&\dots\lambda_{3n}^n\Phi_{k,3n}\end{matrix}\right).\end{eqnarray}

$\lfloor\frac j3\rfloor$ means the floor function of $\frac j3,$ and $\bar j$ means the remainder of $j$ modulo 3.

The $n$-th new solution  $E_1^{[n]},E_2^{[n]}$ after the  n-fold  Darboux transformation of the TH-MB equations will be
\begin{eqnarray}E_1^{[n]}&=&E_1+2i(t_{n-1}^{[n]})_{12},\ \ E_2^{[n]}=E_2+2i(t_{n-1}^{[n]})_{13},
\end{eqnarray}

where $(t_{n-1}^{[n]})_{12}$ is the element at the first row and second column in the matrix of  $t_{n-1}^{[n]}$. The formula for the new solutions $p_1^{[n]},p_2^{[n]},N^{[n]},M_{11}^{[n]},M_{12}^{[n]},M_{21}^{[n]},M_{22}^{[n]}$ is just as the solutions after one fold Darboux transformation $p_1^{[1]},p_2^{[1]},N^{[1]},M_{11}^{[1]},M_{12}^{[1]},M_{21}^{[1]},M_{22}^{[1]}$ in  by changing $\mathbb{T}_1$ to $\mathbb{T}_n$.  So far, we discussed about the determinant construction of n-th Darboux transformation of the TH-MB equations. As an application of these transformations  of the TH-MB equations, soliton solutions and positon solutions will be constructed in the next section.
Particularly we use the 2-fold Darboux transformation to generate two-soliton solutions and positon solutions of the TH-MB equation.

\begin{figure}[h!]
\centering
\raisebox{0.85in}{($|E_1^{[1]}|^2$)}\includegraphics[scale=0.29]{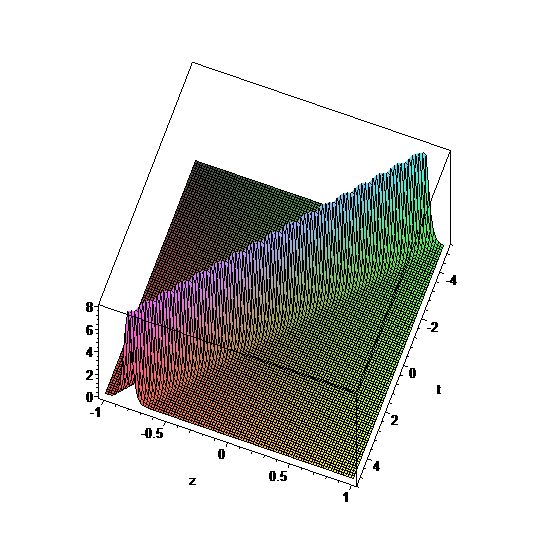}
 \caption{One solition solution $E_1^{[1]}=E_2^{[1]}$  of the TH-MB equations when $,\epsilon=1,\beta=1,\omega=1,\alpha_1=1,\beta_1=2$.}\label{1solitonTH-MB}
\end{figure}

We assume trivial seed solutions as $E_1=0;E_2=0;p_1:=0;p_2:=0;N:=1;M_{11}:=1;M_{12}:=0;M_{21}:=0;M_{22}:=1,$ then the linear equations lead to eigenfunctions
\begin{eqnarray}\label{eigenfunction}
\Phi_1&=&\exp(\frac1{2\lambda+\omega}zi-\lambda ti),\\ \label{eigenfunction3}
\Phi_2&=&\Phi_3=\exp(\frac{(-16\epsilon\lambda^4-8\epsilon\lambda^3\omega+4\beta\lambda^3+2\beta\lambda^2\omega+1)zi}{2\lambda+\omega}+\lambda ti).\end{eqnarray}
 Substituting these two eigenfunctions into the one-fold Darboux transformation eq.\eqref{1darbouxE}and  choosing $\lambda=\alpha_1+\beta_1 i$,  then the following  solition solutions are obtained:

\begin{eqnarray*}E_1^{[1]}=E_2^{[1]}&=&-4\beta_1\frac{e^{\frac{2A}{(2\alpha_1+2i\beta_1+\omega)(-2\alpha_1+2i\beta_1-\omega)}}}
{e^{\frac{16z\epsilon\alpha_1^2\beta_1(12\alpha_1^2+12\omega\alpha_1+3\omega^2+8\beta_1^2)}{(2\alpha_1+2i\beta_1+\omega)(-2\alpha_1+2i\beta_1-\omega)}}+2e^{
\frac{4\beta_1B}{(2\alpha_1+2i\beta_1+\omega)(-2\alpha_1+2i\beta_1-\omega)}}},
\end{eqnarray*}
where
\begin{eqnarray*}A:&=&4t\beta_1^3+48z\epsilon\alpha_1^4\beta_1+32z\epsilon\alpha_1^2\beta_1^3+8z\beta\alpha_1^3\beta_1+8z\beta\beta_1^3\alpha_1\\
&&+4z\epsilon\omega^2\beta_1^3+4t\omega\alpha_1\beta_1+t\alpha_1\omega^2i-16iz\epsilon\alpha_1^5-4iz\beta\beta_1^4+4iz\beta\alpha_1^4\\
&&+4it\omega\alpha_1^2+4it\alpha_1\beta_1^2+48z\epsilon\omega\alpha_1^3\beta_1+12z\epsilon\omega^2\alpha_1^2\beta_1+8z\beta\omega\alpha_1^2\beta_1\\
&&+2z\beta\omega^2\alpha_1\beta_1+16z\epsilon\omega\alpha_1\beta_1^3+16z\epsilon\beta_1^5+4t\alpha_1^2\beta_1+t\beta_1\omega^2+4it\alpha_1^3\\
&&+z\beta\omega^2\alpha_1^2i+32iz\epsilon\alpha_1^3\beta_1^2+48iz\epsilon\beta_1^4\alpha_1+4iz\beta\alpha_1^3\omega-16iz\epsilon\alpha_1^4\omega\\
&&-4iz\epsilon\omega^2\alpha_1^3-iz\beta\omega^2\beta_1^2-4iz\beta\alpha_1\beta_1^2\omega+12iz\epsilon\omega^2\alpha_1\beta_1^2+48iz\epsilon\alpha_1^2\beta_1^2\omega,
\end{eqnarray*}
\begin{eqnarray*}B:&=&16z\epsilon\beta_1^4+4t\alpha_1^2+8z\beta\omega\alpha_1^2+2z\beta\omega^2\alpha_1+4t\omega\alpha_1+16z\epsilon\omega\alpha_1\beta_1^2\\
&&+4z\epsilon\omega^2\beta_1^2+4t\beta_1^2+8z\beta\beta_1^2\alpha_1+t\omega^2+8z\beta\alpha_1^3.
\end{eqnarray*}
 Similarly, substituting the eigenfunctions eq.\eqref{eigenfunction}-eq.\eqref{eigenfunction3} into the one-fold Darboux transformation eq.\eqref{1darbouxE}, and  taking $\alpha=2,\beta=-1$, then the one-solition solutions of the classical TH-MB equations can be obtained whose evolution is given in Fig.\ref{1solitonTH-MB}, which clearly indicates that $E_1=E_2$ are bright solitons because their waves are above the flat non-vanishing plane.
The other solutions keep initial values as following
\begin{eqnarray}N^{[1]} = 1, M_{11}^{[1]} = 1, M_{12}^{[1]} = 0, M_{21}^{[1]} = 0, M_{22}^{[1]} = 1, p_1^{[1]} = 0, p_
2^{[1]} = 0.\end{eqnarray}

\begin{figure}[h!]
\centering
\raisebox{0.85in}{($|E_1^{[2]}|^2$)}\includegraphics[scale=0.25]{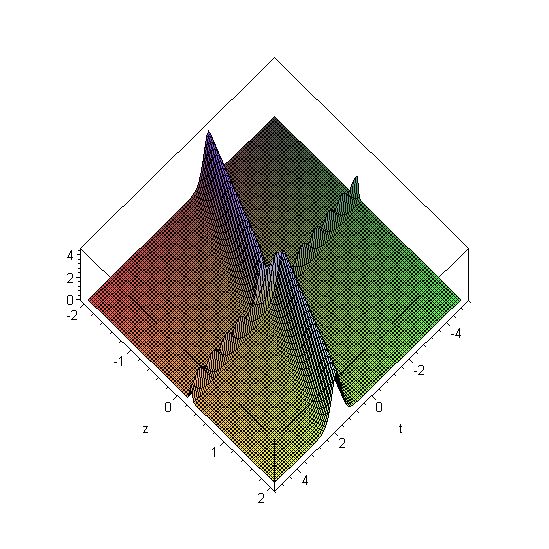}
\hskip 0.03cm
\raisebox{0.85in}{}\includegraphics[scale=0.18]{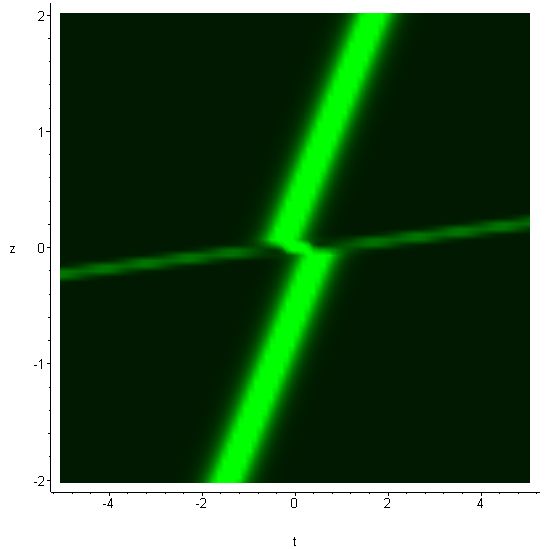}
 \caption{Two solition solution $E_1^{[2]}=E_2^{[2]}$  of the TH-MB equations when $\alpha_1=1,\beta_1=1.5,\alpha_2=1.5,\beta_2=1,\omega=1,\epsilon=1,\beta=1.$}\label{2solitonTH-MB}
\end{figure}

After considering the two-fold Darboux transformation and supposing $\alpha_1=1,\beta_1=1.5,\alpha_2=1.5,\beta_2=1,\omega=1,\epsilon=1,\beta=1,$ two-soliton solutions $E_1^{[2]},E_2^{[2]}$ are as Fig. \ref{2solitonTH-MB}.

For the above two-soliton solutions, if the second spectral parameter $\lambda_2$ is close to the first spectral parameter $\lambda_1$, doing the Taylor expansion of wave function to first order up to $\lambda_1$ will lead to a new kind of solutions which is called positon solutions. Firstly, following four linear eigenfunctions will be got which will be used to construct the second Darboux transformation and  to generate the positon solutions,

\begin{eqnarray}
\Phi_{1,1}&=&\exp(\frac1{2\lambda_1+\omega}zi-\lambda_1 ti),\ \Phi_{4,1}=\exp(\frac1{2\lambda_2+\omega}zi-\lambda_2 ti),\\
\Phi_{2,1}&=&\Phi_{3,1}=\exp(\frac{(-16\epsilon\lambda_1^4-8\epsilon\lambda_1^3\omega+4\beta\lambda_1^3+2\beta\lambda_1^2\omega+1)zi}{2\lambda_1+\omega}+\lambda_1 ti),\\
\Phi_{5,1}&=&\Phi_{6,1}=\exp(\frac{(-16\epsilon\lambda_2^4-8\epsilon\lambda_2^3\omega+4\beta\lambda_2^3+2\beta\lambda_2^2\omega+1)zi}{2\lambda_2+\omega}+\lambda_2 ti).\end{eqnarray}

\begin{figure}[h!]
\centering
\raisebox{0.85in}{($|E_1|^2$)}\includegraphics[scale=0.25]{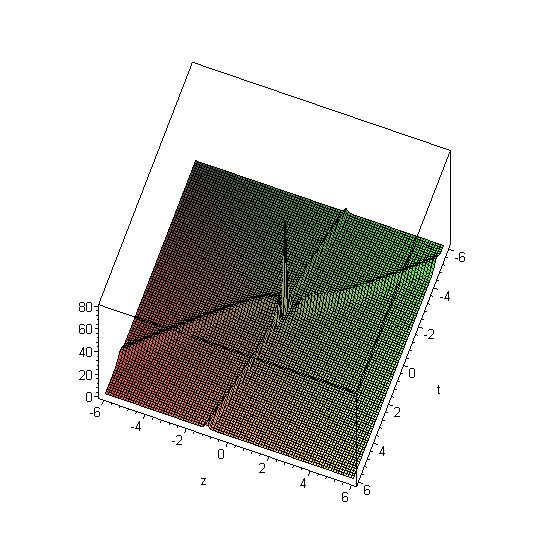}
\hskip 0.03cm
\raisebox{0.85in}{($|E_2|^2$)}\raisebox{-0.1cm}{\includegraphics[scale=0.25]{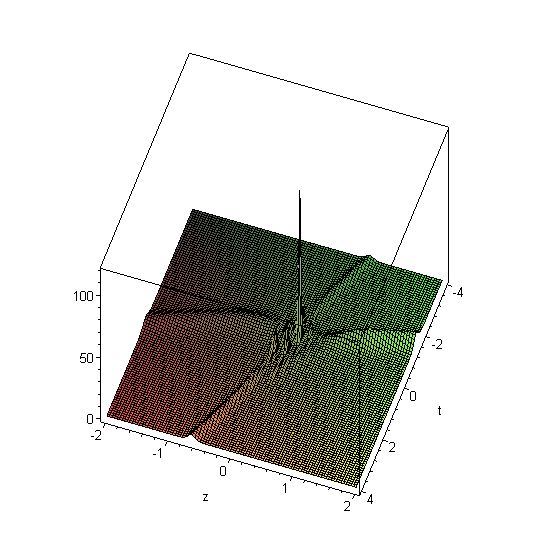}}
 \caption{One positon solution $(E_1,E_2)$  of the TH-MB equations when $\alpha_1=1,\beta_1=1.5,\omega=1,\epsilon=1,\beta=2,\omega=1.5.$}\label{Hpositon}
\end{figure}

Now we take $\lambda_2=\lambda_1+\epsilon(1+i)$ and using the  Taylor expansion of wave function $\Phi_{4,1},\Phi_{5,1},\Phi_{6,1}$ up to first order of $\epsilon$ in terms of $\lambda_1$.  On substitution of these manipulations into the second Darboux transformation discussed in the last section will help us to derive positon solutions.
The pictorial representation of positon solutions of the TH-MB equations, i.e. the case when $\alpha_1=1,\beta_1=1.5,\omega=1,\epsilon=1,\beta=2,\omega=1.5$ is shown in Fig.\ref{Hpositon}.

\begin{figure}[h!]
\centering
\raisebox{0.85in}{($|E_1|^2$)}\includegraphics[scale=0.25]{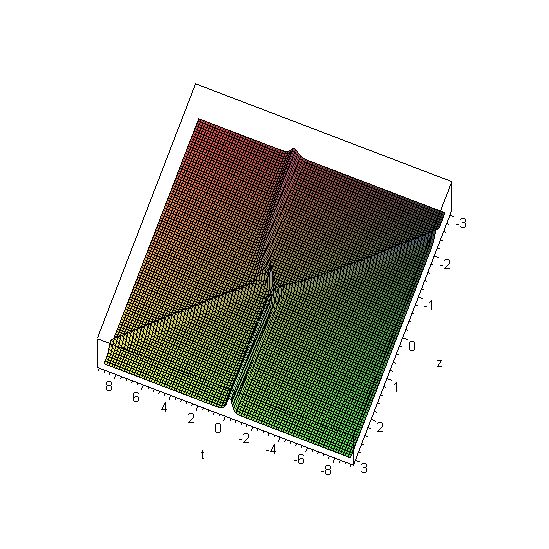}
\hskip 0.03cm
\raisebox{0.85in}{($|E_2|^2$)}\raisebox{-0.1cm}{\includegraphics[scale=0.25]{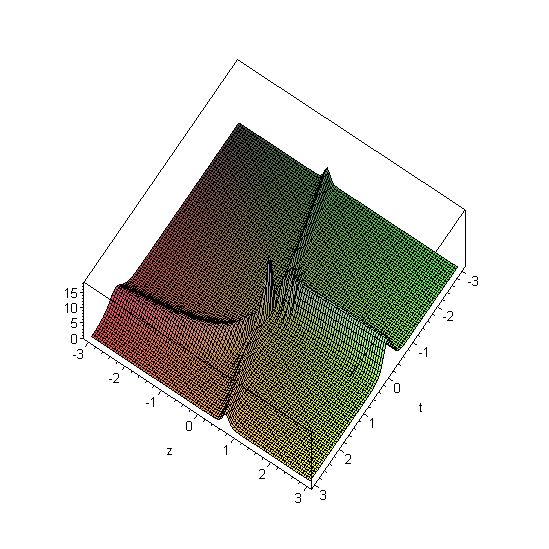}}
 \caption{One positon solution $(E_1,E_2)$  of the TMKDV-MB equations when $\alpha_1=1,\beta_1=1.5,\omega=1,\epsilon=1,\beta=0,\omega=1.5.$}\label{01positon}
\end{figure}

When $\alpha_1=1,\beta_1=1.5,\omega=1,\epsilon=1,\beta=0,\omega=1.5$, i.e. the case of two-component complex modified Korteweg-de Vries(TCMKdV)-MB equations,
the picture of positon solutions of the TCMKdV-MB equations is plotted in Fig.\ref{01positon}.

Comparing positons with two solitons, one can find positons are long-range analogues of solitons and slowly decreasing, oscillating solutions.
Though each branch of the positon will separate finally during their propagating,
the separation becomes slower and slower because $\lambda_2$ infinitely approach to $\lambda_1$.

 In a similar way, using the higher order Darboux transformation, one can also generate higher-order positon solutions which will not be included here.

The authors would like to thank Professor Jingsong He for helpful discussions.




\end{document}